\documentclass[aps,prd,preprintnumbers,showpacs]{revtex4}
\usepackage[dvips]{graphicx}

\begin{document}

%
%

\eprint{Nisho-2-2021}
\title{Radiation Production by Axion in Space between Two Flat Conductors }
\author{Aiichi Iwazaki}
\affiliation{International Economics and Politics, Nishogakusha University,\\ 
6-16 3-bantyo Chiyoda Tokyo 102-8336, Japan.}   
\date{July. 12, 2021}
\begin{abstract}
We have shown a new production mechanism of radiations produced by dark matter axion $a(t)$
under strong magnetic field $B$. The axion generates oscillating electric current in the surface of conductor.
The current is extremely enhanced such as $J_c=(m_a\delta_e)^{-2}J_a$ 
compared with vacuum current $J_a\equiv g_{a\gamma\gamma}\partial_t a(t)B$; $\delta_e$ 
is skin depth
and axion mass $m_a$. We have shown that sufficiently strong radiation 
is emitted by cylindrical super ( normal ) conductor to be easily detected.
Then, we naively expect that such strong radiation also arises in resonant cavity experiment.
But we show that 
the expectation is wrong. The radiation generated in the cavity is
just the one given in the standard estimation, even if we consider the
enhanced current in the conductor of the cavity. 
This is because the radiation in the cavity is standing wave, while 
the radiation emitted from the cylindrical conductor is outgoing wave.  
For simplicity we consider radiation arising in the space between two parallel flat conductors.

\end{abstract}
\hspace*{0.3cm}

\hspace*{1cm}

\maketitle

\section{introduction}
The axion\cite{axion, Wil} is one of the most important candidates of the dark matter.
Especially, it appears that it stands on the top of the candidates because 
QCD axion gives natural solution of strong CP problem and other candidates of the dark matter 
e.g. WINPs ( weak interacting massive particles )  
have not been detected in spite of extensive searches for long terms.  

The axion has recently been searched with various ways.
Most of the experiments\cite{admx, carrack, haystac, abracadabra, organ, madmax, brass, cast, sumico}  exploit the Primakoff effect. That is, 
they detect photon directly produced by the dark matter axion $a(t)$ under strong magnetic field $B$.
But, the axion photon coupling $g_{a\gamma\gamma}a(t)\sim 10^{-19}$ is so tiny that the direct photon production is
hard to be detected. So we need, for instance, very sensitive radio receiver using 
fine technology such as SQUIDs.
( The dark matter axion in our galaxy has very small momentum $\vec{p}_a\sim 10^{-3}m_a$ so that we may neglect the momentum dependence of
the axion field $a(\vec{p}_a\cdot \vec{x},t)\simeq a(t)$ in terrestrial experiments. )

In the standard resonant cavity experiment, the radiation energy is estimated\cite{sikivie,will} supposing
that the axion-photon conversion arises in the cavity with
the assumption of perfect conductor. The energy density is estimated such as $\sim (g_{a\gamma\gamma}B/m_a)^2\rho_s Q^2$ with
the dark matter axion density $\rho_a$ and $Q$-value of the cavity.
We may understand that they are produced by a fictitious oscillating electric current $J_a\sim g_{a\gamma\gamma}\partial_t a(t)B$ of the axion 
induced under external magnetic field $B$.

\vspace{0.1cm}
On the other hand, we have shown in recent papers\cite{1iwazaki}
that radiations are emitted from cylindrical conductor by oscillating electric current $\sigma E_a$ with electrical conductivity $\sigma$.
The current is induced in the conductor by the oscillating electric field $E_a=-g_{a\gamma\gamma}a(t)B$
under external magnetic field $B$.
The current flows only in the surface of the conductors due to the skin effect.
The amount of the current density
is enhanced by the factor $1/(m_a\delta)^2\sim 10^8(10^{-5}\mbox{eV}/m_a)(\sigma/10^3\rm eV)$, 
compared by the fictitious current $J_a$ of the axion.
We have shown that sufficiently large amount of the radiations are emitted from the conductor
to be easily detected with existing radio telescopes. 

\vspace{0.1cm}

According to our mechanism of radiation production,
we naively expect that such large amount of radiation arises even in resonant cavity.
Namely, because the oscillating electric field $E_a$ is produced 
inside the conductor of the cavity, the enhanced electric current flows in the conductors and emit strong electromagnetic fields.
But we show that this is not true because the radiations are standing wave in the cavity.

Even if strong radiation is initially emitted by the enhanced electric current, it reflects from the other side of the cavity
and goes back to the original side of the cavity. Then, it penetrates the conductor of the cavity
and makes the electric current diminished. As a consequence, 
the standing wave of the radiation is not strong. Its strength is determined by the
radiation converted from the axion. That is, the strength of the standing wave radiation in the cavity is the same as
$E_a\sim g_{a\gamma\gamma}a(t)B$ produced by the axion in vacuum, unless the radiation is resonantly enhanced.

\vspace{0.1cm}
The situation is similar to the case that electromagnetic radiation enters a flat conductor from outside and
reflects from it, going out. It is the scattering of the radiation by the conductor.
The radiation is stationary wave; superposition of incoming and outgoing waves.
We know that the radiation penetrates in the surface of the conductor to the skin depth.
Thus, electric current is induced in the surface. The magnitude of the current depends on
the electrical conductivity. It implies that we have various magnitude of the currents,
although the incoming radiation is identical.  
The situation is different to the case of outgoing radiation emitted by AC current. 
It's strength is determined by the amount of the AC current.

\vspace{0.1cm}
In this letter, for simplicity, instead of resonant cavity,
we discuss radiation in the space between two flat conductors ( slabs ) put parallel to each other.
The radiation we discuss is standing wave generated by the axion. 
Even if we take account of the enhanced electric current
in the conductors,
we see that the strength of the radiation is identical to the electric field $E_a\sim g_{a\gamma\gamma}a(t)B$ 
converted from the axion. It is the same as radiations previously shown in the cavity experiments.  
It does not depend on the electrical conductivity of the conductors.
By tuning the spacing $l=\pi/m_a$ between two flat conductors, we have
resonantly enhanced radiation. This is similar to the resonance in the cavity.

\section{radiation between two flat normal conductors}
We discuss radiation in the space between two parallel flat normal conductors.
First, we explain the axion photon coupling.

The axion $a(\vec{x},t)$ couples with both electric $\vec{E}$ and magnetic fields $\vec{B}$ in the following,

\begin{equation}
\label{L}
L_{aEB}= g_{\gamma}\alpha \frac{a(\vec{x},t)\vec{E}\cdot\vec{B}}{f_a\pi}\equiv g_{a\gamma\gamma} a(\vec{x},t)\vec{E}\cdot \vec{B}
\end{equation}
with the decay constant $f_a$ of the axion
and the fine structure constant $\alpha\simeq 1/137$,   
where the numerical constant $g_{\gamma}$ depends on axion models; typically it is of the order of one.
The standard notation $g_{a\gamma\gamma}$ is such that $g_{a\gamma\gamma}=g_{\gamma}\alpha/f_a\pi\simeq 0.14(m_a/\rm GeV^2)$
for DFSZ model\cite{dfsz} and $g_{a\gamma\gamma}\simeq -0.39(m_a/\rm GeV^2)$ for KSVZ model\cite{ksvz}.
In other words, $g_{\gamma}\simeq 0.37$ for DFSZ and $g_{\gamma}\simeq -0.96$ for KSVZ.
The axion decay constant $f_a$ is related with the axion mass $m_a$ in the QCD axion; $m_af_a\simeq 6\times 10^{-6}\rm eV\times 10^{12}$GeV.

\vspace{0.1cm}
We show that
the coupling parameter $g_{a\gamma\gamma} a(t)$ in the Lagrangian eq(\ref{L}) is extremely small for the dark matter axion $a(t)$.
We note that the energy density of the dark matter axion taken by time average is given by

\begin{equation}
\rho_a= \frac{1}{2}\overline{(\dot{a}^2+(\vec{\partial} a)^2+m_a^2a^2)}\simeq \frac{m_a^2a_0^2}{2}
\end{equation}
where $a(t)=a_0\cos(t\sqrt{m_a^2+p_a^2} )\simeq a_0\cos(m_a t)$, because the velocity $p_a/m_a$ of the axion is about $10^{-3}$ in our galaxy.
The local energy density $\rho$ of dark matter in our galaxy is supposed such as $\rho\simeq 0.3\rm GeV\,\rm cm^{-3}\simeq 2.4\times 10^{-42}\rm GeV^4$.

Assuming that the density is equal to that of the dark matter axion, i.e. $\rho=\rho_a$,
we find extremely small parameter $g_{a\gamma\gamma}a\sim 10^{-19}$.
The energy density also gives the large number density of
the axions $\rho_a/m_a\sim 10^{15}\mbox{cm}^{-3}(10^{-6}\mbox{eV}/m_a)$, which give rise to their coherence. 
This allows us to treat the axions as the classical axion field $a(t)\propto \cos(m_at)$ spatially smoothly varying. Anyway, 
we find that the parameter $g_{a\gamma\gamma} a(t)$ is extremely small.

\vspace{0.2cm}
The interaction term in eq(\ref{L}) between axion and electromagnetic field slightly modifies Maxwell equations in vacuum,

\begin{eqnarray}
\label{modified}
\vec{\partial}\cdot\vec{E}+g_{a\gamma\gamma}\vec{\partial}\cdot(a(\vec{x},t)\vec{B})&=0&, \quad 
\vec{\partial}\times \Big(\vec{B}-g_{a\gamma\gamma}a(\vec{x},t)\vec{E}\Big)-
\partial_t\Big(\vec{E}+g_{a\gamma\gamma}a(\vec{x},t)\vec{B}\Big)=0,  \nonumber  \\
\vec{\partial}\cdot\vec{B}&=0&, \quad \vec{\partial}\times \vec{E}+\partial_t \vec{B}=0.
\end{eqnarray}
From the equations, we approximately obtain the electric field $\vec{E}$
generated by the axion $a$ under the static background magnetic field $\vec{B}(\vec{x})$,

\begin{equation}
\label{E}
\vec{E}_a(r,t)=-g_{a\gamma\gamma}a(\vec{x},t)\vec{B}(\vec{x}),
\end{equation}
assuming the parameter $g_{a\gamma\gamma} a$ extremely small.
This is the electric field in vacuum induced by the dark matter axion $a(\vec{x},t)=a_0\cos(\omega_a t-\vec{p}_a\cdot \vec{x})\simeq a_0\cos(m_at)$ 
with $\omega_a=\sqrt{m_a^2+p_a^2}\simeq m_a$.
We note that the magnetic field configuration is arbitrary.

In addition to the electric field, we have propagating electromagnetic fields $\delta \vec{E}$ and $\delta \vec{B}$ in vacuum.
They are produced by electric current in conductors induced by the axion.

\vspace{0.1cm}
In order to obtain electromagnetic fields between two parallel flat conductors, 
we suppose conductors with permeability $\mu$, 
dielectric permittivity $\epsilon$ and electrical conductivity $\sigma$. 
The configurations of the conductors are in the following. The conductors with flat surfaces are put parallel to each other.
They occupy the regions $x \ge 0$ and $x \le -l$. Their surfaces are present at $x=0$ and $x=-l$, respectively. They
are uniform in $y$ and $z$ directions.
We impose magnetic field $\vec{B}=(0,0,B_0)$ parallel to the surfaces of the conductors. 
We denote the magnetic
field inside as $\vec{B}_{in}$. 
Because the component of the field $\vec{H}_{in}=\vec{B}_{in}/\mu$ parallel to the surface ( at $x=0$ and $x=-l$ ) is continuous at the surface,
the magnetic field $\vec{B}_{in}$ is obtained such that $\vec{B}_{in}=\vec{B}\mu/\mu_0$ where we denote the vacuum permeability $\mu_0$ ( $\mu_0=1$ in natural unit ).
Hereafter, for simplicity we assume the values $\mu=1$ and $\epsilon=1$. 
Thus,  $\vec{B}_{in}(x=0,-l)=\vec{B}(x=0,-l)$.

The electromagnetic fields in the conductor are described by the modified Maxwell equations,

\begin{eqnarray}
\label{1metal}
\vec{\partial}\cdot\vec{E}_{in}+g_{a\gamma\gamma}\vec{\partial}\cdot(a(\vec{x},t)\vec{B}_{in})&=0&, \quad 
\vec{\partial}\times \Big(\vec{B}_{in}-g_{a\gamma\gamma} a(\vec{x},t)\vec{E}_{in}\Big)-
\partial_t\Big(\vec{E}_{in}+g_{a\gamma\gamma} a(\vec{x},t)\vec{B}_{in}\Big)=\vec{J}_e,  \nonumber  \\
\vec{\partial}\cdot\vec{B}_{in}&=0&, \quad \vec{\partial}\times \vec{E}_{in}+\partial_t \vec{B}_{in}=0.
\end{eqnarray}

In eq(\ref{1metal}) we have included the current $\vec{J}_e=\sigma\vec{E}_{in}$ induced by electric field $\vec{E}_{in}$, but have neglected external current
generating the background static magnetic field $\vec{B}$. 

When we neglect axion contribution, we obtain magnetic field $\vec{B}^0_{in}=\vec{B}$ uniform inside the conductor.
Obviously, there is no electric field inside the conductor.
When we take into account the axion contribution, the oscillating electric field is induced. Naively we expect that the electric field is given such that 
$\vec{E}_a=-g_{a\gamma\gamma} a(t)\mu\vec{B}$. But, this is not correct as we show below.
 
Assuming the parameter $g_{a\gamma\gamma} a(\vec{x},t)$ small and noting that the electric field is of the order of  $g_{a\gamma\gamma} a(\vec{x},t)$,
we derive the equations,

\begin{equation}
\label{1max}
\vec{\partial}\cdot \vec{E}_{in}=0, \quad \vec{\partial}\times \vec{B}_{in}=\vec{J}_e+\partial_t(\vec{E}_{in}-\vec{E}_a), \quad
\vec{\partial}\cdot \vec{B}_{in}=0, \quad \mbox{and} \quad \vec{\partial}\times \vec{E}_{in}+\partial_t\vec{B}_{in}=0,
\end{equation}
where we have used the relation $\vec{\partial}\times \vec{E}_a=0$ because $\vec{\partial}\times \vec{B}=0$ inside the conductor.

Using the Ohm's law $\vec{J}_e=\sigma\vec{E}_{in}$ in eqs(\ref{1max}), we derive 
the equation for $\vec{E}_{in}$,

\begin{equation}
\label{19}
(\vec{\partial}^2-\partial_t^2)\vec{E}_{in}=\sigma\partial_t\vec{E}_{in}-\partial_t^2\vec{E}_a
\end{equation}
where we note that $\vec{E}_a\propto \cos(m_at)$. Then, we find the solution for $x>0$ ( later we discuss the solution for $x<-l$ ),

\begin{equation}
\label{21}
\vec{E}_{in}\simeq \vec{E}_0\exp(-\frac{x}{\delta_e})\cos(\omega t-\frac{x}{\delta_e}+\delta)+\frac{1}{\sigma}\partial_t \vec{E}_a,
\end{equation} 
with arbitrary field strength $\vec{E}_0$, and frequency $\omega$. The skin depth $\delta_e$ is given by $\delta_e=\sqrt{2/\mu\sigma\omega}$.
In the derivation, we have neglected the term $\partial_t^2\vec{E}_{in}$ in the left hand side of eq(\ref{19}), 
which is much smaller than the term $\sigma\partial_t\vec{E}_{in}$
in the right hand side.
Namely, we have used the inequality $\sigma \gg \omega \, ( \sim m_a) $ 
because the electric conductivity $\sigma \sim 10^4$eV of copper is much larger than 
the axion mass $m_a=10^{-6}$eV $\sim 10^{-3}$eV under consideration.

The strength $\vec{E}_0$ ( frequency $\omega$ ) is determined by the boundary condition at the surface $x=0$.
In our case it is determined by the electric field induced outside the conductor, i.e. electric field in the vacuum. It is 
just $\vec{E}_a=-g_{a\gamma\gamma} a(t)\vec{B}$.
Thus, $\vec{E}_0=(0,0,-g_{a\gamma\gamma} a_0B_0)$ and $\omega=m_a$.
Then, it induces large electric current $\sigma E_a$ which emits radiation from the conductor present in the region $x<0$.

In addition to the electric field, we need to take into account the electric field of the radiation penetrating from
outside the conductors. The radiation is the reflected one from the conductor in the other side $x<-l$.
A superposition of
outgoing and incoming radiations is a standing wave in the space between two flat conductors. The wave
penetrates the conductor so that we have an additional electric field.

Therefore, we have electric fields in the conductor,

\begin{eqnarray}
\label{ele}
\vec{E}_{in}&\simeq& \vec{E}_0\exp(-\frac{x}{\delta_e})\cos(\omega t-\frac{x}{\delta_e})
+\vec{E}_1\exp(-\frac{x}{\delta_e})\Big(\cos(\omega t-\frac{x}{\delta_e})+A\sin(\omega t-\frac{x}{\delta_e})\Big)+\frac{1}{\sigma}\partial_t \vec{E}_a, \\
&\equiv&\vec{E}_0\exp(-\frac{x}{\delta_e})\cos(\omega t-\frac{x}{\delta_e})+\delta \vec{E}(\mbox{inside}) \\
 \delta \vec{E}(\mbox{inside})&=&\vec{E}_1\exp(-\frac{x}{\delta_e})\Big(\cos(\omega t-\frac{x}{\delta_e})
 +A\sin(\omega t-\frac{x}{\delta_e})\Big)+\frac{1}{\sigma}\partial_t \vec{E}_a,
\end{eqnarray} 
with constants $\vec{E}_1=(0,0,E_1)$ and $A$ determined by radiations outside the conductor. 
Actually, boundary conditions at $x=0$ and $x=-l$ leads to the constants.

\vspace{0.1cm}
The first and second terms in the solution $\vec{E}_{in}$ in eq(\ref{ele}) 
represents oscillating electric fields with the skin depth $\delta_e$, while the last term represents
the oscillating electric field $\frac{1}{\sigma}\partial_t \vec{E}_a$ present uniformly inside the conductor. 
This term is absent when we consider conductor with finite length in $z$ direction 
because electric charges on two ends of upper and lower surfaces 
induced by the electric field screen the field $\frac{1}{\sigma}\partial_t \vec{E}_a$.
( We neglect the term hereafter. )  
Although the electric charges on the surfaces oscillate with the frequency $m_a$ 
and produce electric currents\cite{wire} with the skin depth $\delta_e$, the current density 
$\frac{1}{\sigma}\partial_t^2 \vec{E}_a\sim (m_a\delta_e)^2J_a\sim m_a/\sigma J_a$ is much small 
compared with the current $\sigma E_0\sim 1/(m_a\delta_e)^2J_a$ or even $\sqrt{\sigma/m_a}J_a$. 
( The current $\sqrt{\sigma/m_a}J_a$ is derived by $\sigma (E_1+E_0)\sim \sigma m_a\delta_e E_a\sim \sqrt{\sigma/m_a}J_a$ 
for ${m_a\delta_e} \ll 1$, see later .)
Thus, 
we neglect the last term in the discussion below. 

\vspace{0.1cm}
Because we have electric field $\vec{E}_{in}$, the electric current $\vec{J}_e=\sigma \vec{E}_{in}$ flows in the surface of the conductor.
The current produces radiation in the space between two conductors. 
From the Maxwell equation $\vec{\partial}\times \delta \vec{B}_{in}=\vec{J}_e+\partial_t\vec{E}_{in}$,
we obtain the magnetic field $\delta B_y(\mbox{inside})$ generated by the current $\vec{J}_e=\sigma \vec{E}_{in}$,
or by noting $\partial_t \delta B_y=\delta_x E_{in}$,

\begin{eqnarray}
\label{mag}
\delta B_y(\mbox{inside})&=&-\frac{E_0}{\delta_e m_a}\exp\big(-\frac{x}{\delta_e}\big)\Big(\sin(m_a t-\frac{x}{\delta_e})
+\cos(m_a t-\frac{x}{\delta_e})\Big) \nonumber \\
&-&\frac{E_1}{\delta_e m_a}\exp\big(-\frac{x}{\delta_e}\big)\Big(\sin(m_a t-\frac{x}{\delta_e})
+\cos(m_a t-\frac{x}{\delta_e})\Big) \nonumber \\
&-&\frac{AE_1}{\delta_e m_a}\exp\big(-\frac{x}{\delta_e}\big)\Big(\sin(m_a t-\frac{x}{\delta_e})
-\cos(m_a t-\frac{x}{\delta_e})\Big),
\end{eqnarray}
with $E_0=-g_{a\gamma\gamma} a_0B_0$. 
The magnetic field $\delta \vec{B}(\mbox{inside})=(0,\delta B_y,0) $ as well as the electric field $\vec{E}_{in}=(0,0,E_{in})$ are induced by 
the dark matter axion $a(t)$ in the conductor ( $x<0$ ). 

\vspace{0.1cm}
Using the electric and magnetic fields $\delta E_z(\mbox{inside})$ and $\delta B_y(\mbox{inside})$ 
inside the conductor, we can determine the standing wave of electromagnetic field 
in the space between the two conductors. We consider only the mode such that $\delta\vec{B}(\mbox{outside})=(0,\delta B_y,0)$ and 
$\delta \vec{E}(\mbox{outside})=(0,0,\delta E_z)$. Then, the solutions of the Maxwell equations in vacuum are
\begin{eqnarray}
\label{st}
\delta B_y(\mbox{outside})&=&b_1\sin(m_ax+\delta_1)\sin(m_at)+b_2\sin(m_ax+\delta_2)\cos(m_at) \\
\delta E_z(\mbox{outside})&=&-b_1\cos(m_ax+\delta_1)\cos(m_at)+b_2\cos(m_ax+\delta_2)\sin(m_at)
\end{eqnarray}
with 

\begin{eqnarray}
b_1\sin\delta_1&=&\frac{-E_0}{m_a\delta_e}-\frac{E_1(1+A)}{m_a\delta_e}, \quad b_2\sin\delta_2=\frac{-E_0}{m_a\delta_e}-\frac{E_1(1-A)}{m_a\delta_e} \\
-b_1\cos\delta_1&=&E_1, \quad b_2\cos\delta_2=A E_1,
\end{eqnarray}
where we have imposed the boundary condition $\delta B_y(\mbox{outside})=\delta B_y(\mbox{inside})$, and 
$\delta E_z(\mbox{outside})=\delta E_z(\mbox{inside})$ at $x=0$.
Similarly, we also impose the boundary condition at $x=-l$ and obtain

\begin{eqnarray}
b_1\sin(\delta_1-m_al)&=&\frac{E_0}{m_a\delta_e}+\frac{E_1(1+A)}{m_a\delta_e} \quad 
b_2\sin(\delta_2-m_al)=\frac{E_0}{m_a\delta_e}+\frac{E_1(1-A)}{m_a\delta_e} \\
-b_1\cos(\delta_1-m_al)&=&E_1 \quad 
b_2\cos(\delta_2-m_al)= AE_1
\end{eqnarray}

In order to obtain the above results, 
we note that the solutions of the electric and magnetic fields inside the conductor $x \le -l$ are given by

\begin{eqnarray}
\vec{E}_{in}&=& \vec{E}_0\exp\Big(\frac{x+l}{\delta_e}\Big)\cos\big(\omega t+\frac{x+l}{\delta_e}\big)
+\vec{E}_1\exp\Big(\frac{x+l}{\delta_e}\Big)\Big(\cos(\omega t+\frac{x+l}{\delta_e})+A\sin(\omega t+\frac{x+l}{\delta_e})\Big) \nonumber \\
\delta B_y(\mbox{inside})&=& \frac{E_0}{m_a\delta_e}\exp\Big(\frac{x+l}{\delta_e}\Big)
\Big( \sin\big(m_at+\frac{x+l}{\delta_e}\big)+\cos\big(m_at+\frac{x+l}{\delta_e}\big)\Big) \nonumber \\
&+&\frac{E_1}{\delta_e m_a}\exp\big(\frac{x+l}{\delta_e}\big)\Big(\sin(m_a t+\frac{x+l}{\delta_e})
+\cos(m_a t+\frac{x+l}{\delta_e})\Big) \nonumber \\
&+&\frac{AE_1}{\delta_e m_a}\exp\big(\frac{x+l}{\delta_e}\big)\Big(\sin(m_a t+\frac{x+l}{\delta_e})
-\cos(m_a t+\frac{x+l}{\delta_e})\Big) \nonumber \\
&\mbox{for}& \quad  x \le -l 
\end{eqnarray}

\vspace{0.1cm}
Therefore, we find the coefficients, $E_1$ and $A$ as well as the phases $\delta_{1,2}$ by noting $m_a\delta_e \ll 1$ 
( $m_a\delta_e\sim 10^{-4}\sqrt{m_a/10^{-5}\mbox{eV}}\sqrt{10^3\mbox{eV}/\sigma}$ )
such that 

\begin{eqnarray}
\label{coef}
E_1&\simeq& -E_0\Big(1+\frac{1}{2}m_a\delta_e\tan\delta_1\Big), \quad A\simeq \frac{m_a\delta_e\tan\delta_1}{2} \nonumber \\ 
\quad b_1&\simeq &\frac{E_0}{\cos\delta_1}\Big(1+\frac{m_a\delta_e}{2}\tan\delta_1\Big) , 
\quad b_2\simeq -\frac{E_0m_a\delta_e\tan\delta_1}{2\cos\delta_2}
\end{eqnarray}
and the phases satisfying the identical equation,

\begin{equation}
\tan\delta_{1,2}=\frac{\sin(m_al)}{1+\cos(m_al)}.
\end{equation}

Therefore, the coefficients are

\begin{equation}
E_1= -E_0\Big(1+O(m_a\delta_e)\Big), \quad A=O(m_a\delta_e), \quad b_1=\frac{E_0}{\cos\delta_1}\Big(1+O(m_a\delta_e)\Big), \quad b_2=O(m_a\delta_e).
\end{equation}

Therefore,
we find that the standing waves in eq(\ref{st}) of the electromagnetic fields in the space between two parallel flat conductors are
given in the limit $m_a\delta_e \to 0$, that is, $b_1=E_0/\cos\delta_1$ and $b_2=0$ by,
\begin{eqnarray}
\label{field}
\delta B_y(\mbox{outside})&=&\frac{E_0}{\cos\delta_1}\sin(m_ax+\delta_1)\sin(m_at) \nonumber \\
\delta E_z(\mbox{outside})&=&-\frac{E_0}{\cos\delta_1}\cos(m_ax+\delta_1)\cos(m_at).
\end{eqnarray}

We note that the field strength $E_0=-g_{a\gamma\gamma}a_0B_0$ 
is the same as the electric field $E_a=-g_{a\gamma\gamma}a_0\cos(m_at)B_0$ 
converted from the axion in the vacuum. ( $\cos\delta_1$ is of the order of $1$ .) 
Namely, the strengths of the standing waves in the space between two flat conductors are identical to the electric field
generated by the axion in vacuum. 

\vspace{0.1cm}
We would like to make a comment that when we take into account only the first term $E_a$ of $E_{in}$ in eq(\ref{ele}) and eq(\ref{mag})
to determine $\delta B_y(\mbox{outside})$, the electric field $E_a$
generates the enhanced electric current $\sigma E_a\propto J_a/(m_a\delta_e)^2$ with $J_a\sim g_{a\gamma\gamma}\partial_t a(t)B$ and
the radiation between two flat conductors has a enhancement
factor $1/(m_a\delta_e)$. Namely, 
the magnetic field outside the conductor becomes such that $\delta B_y(\mbox{outside})\propto E_a/(m_a\delta_e)$.
The radiation field satisfies the boundary condition $\delta B_y(\mbox{outside})=\delta B_y(\mbox{inside})$,
but, the corresponding electric field does not satisfy 
the boundary condition $\delta E_z(\mbox{outside})=\delta E_z(\mbox{inside})$ at $x=0$.
In other words, the magnetic field $\delta B_y(\rm outside )$ is determined by the current in the surface. 

The boundary condition on $\delta B_y$ implies that the radiation with the enhanced magnetic field $\delta B_y(\mbox{outside})$ is emitted by the
the enhanced electric current $\sigma E_a$.
That is, the field represents only the wave emitted by the enhanced current so that it does not
satisfy the boundary condition on the electric field. For the standing wave, we need to take into account
the back reaction of the field penetrating the conductor.  That is, we should consider
the second term of $E_{in}$ in eq(\ref{ele}) which represents the electric field coming into the conductor $x<0$ from the outside $x>0$.
Then, the resultant standing wave satisfies the boundary condition on the electric field as well as magnetic field,
as we have shown above. 

\vspace{0.1cm}
Obviously, the field in eq(\ref{field}) does not
depend on the electrical conductivity $\sigma$ of the conductor. It only depends on the electric field $E_a=-g_{a\gamma\gamma}a(t)B_0$
naively produced by the axion. The result has been obtained assuming perfect conductor. That is, $\sigma\to \infty$
so that $m_a\delta_e \to 0$.  We note that 
the previous estimation of the radiation in cavity was obtained\cite{will} 
assuming perfect conductor. Therefore, we have confirmed that the estimation is
valid even if we take into account the electric currents inside the conductors induced by the dark matter axion. 

\vspace{0.1cm}
Contrary to the standing wave, when we consider outgoing radiation emitted by AC electric current, the strength of
the radiation is determined by the AC current. We do not need to consider the back reaction of the radiation on the AC current.
For instance, 
we have AC voltage generating by the axion under external magnetic field imposed parallel on the surface of 
a single flat conductor. Then, AC current flows in the conductor and emits outgoing radiation.
It never comes back to the conductor. As is well known,
the strength of the radiation is determined by the AC current, which is given by the AC voltage generated by the axion.
This is different to the situation of 
standing wave between the conductors put parallel to each other.
The strength of the standing wave is determined
by the AC voltage, not AC current. 
The point we have mentioned is very important for the determination of the power of radiations caused by the axion
under external magnetic field.

\vspace{0.2cm}
We proceed to estimate the time average of the electromagnetic energy density,

\begin{equation}
U=\frac{\overline{(\delta E_z^2)}+\overline{(\delta B_y)^2}}{2}=\frac{1}{4}\Big(g_{a\gamma\gamma}a_0B_0\Big)^2
\Big(\frac{1}{\cos\delta}\Big)^2
\end{equation}
with $\delta\equiv \delta_{1,2}$.

\vspace{0.1cm}
The energy density $U$ is divergent when $\cos\delta \to 0$. Because $1/\cos\delta \to \frac{2}{(\pi/l-m_a)l}$ in the limit $l \to \pi/m_a$,
we obtain

\begin{equation}
U\simeq  \frac{1}{4}\Big(g_{a\gamma\gamma}a_0B_0\Big)^2
\Big(\frac{2}{m_a-\pi/l}\Big)^2\Big(\frac{1}{l}\Big)^2=\Big(g_{a\gamma\gamma}B_0\Big)^2\Big(\frac{2\rho_a}{m_a^2}\Big)
\Big(\frac{1}{m_a-\pi/l}\Big)^2\Big(\frac{1}{l}\Big)^2
\end{equation}
where $\rho_a$ denotes the energy density of the dark matter axion.

\vspace{0.1cm}

Up to now, we have assumed that the energy $\omega_a$ of the axion is given by $\omega_a=\sqrt{p_a^2+m_a^2} \simeq m_a$.
In the real situation, the energy $\omega_a$ of the dark matter axion is supposed to have the distribution $f(\omega_a)$ such that 

\begin{equation}
f(\omega_a)=f_0\exp\Big(-\frac{\omega_a-m_a}{\epsilon}\Big)\sqrt{\omega_a-m_a}\,\theta(\omega_a-m_a)
  \quad \mbox{with} \quad \epsilon =m_a\langle v^2 \rangle/3
\end{equation}
with $\int_{m_a}^{\infty} d\omega_a f(\omega_a)=1$ ( $f_0=2\epsilon^{-3/2}/\sqrt{\pi} $ ), 
where $\langle v^2 \rangle$ is the average of  the square of axion velocity $v$ near the earth.
We assume that $\langle v^2 \rangle \simeq 10^{-6}$.

Then, the energy density is given by

\begin{equation}
U=\int \Big(g_{a\gamma\gamma}B_0\Big)^2\Big(\frac{2\rho_a}{\omega_a^2}\Big)
\Big(\frac{1}{\omega_a-\pi/l}\Big)^2\Big(\frac{1}{l}\Big)^2 f(\omega_a)d\omega_a
\end{equation}

The integral is divergent because of the factor $1/(\omega_a-\frac{\pi}{l})^2$. 
The divergence is regularized\cite{will,kim} in the standard manner by taking account of dissipation of field energy such that

\begin{eqnarray}
U&=&\int \Big(g_{a\gamma\gamma}B_0\Big)^2\Big(\frac{2\rho_a}{\omega_a^2}\Big)
\Big(\frac{1}{(\omega_a-\pi/l)^2+\frac{\omega_a^2}{4Q^2}}\Big)\Big(\frac{1}{l}\Big)^2 f(\omega_a)d\omega_a \\
&\simeq& 4\Big(g_{a\gamma\gamma}B_0\Big)^2\Big(\frac{2\rho_a}{m_a^2}\Big)
\Big(\frac{Q^2}{m_a^2}\Big)\Big(\frac{1}{l}\Big)^2\simeq \frac{8}{\pi^2}\Big(\frac{g_{a\gamma\gamma}B_0}{m_a}\Big)^2\rho_aQ^2
\end{eqnarray}
with $Q$-value of the cavity and $l=\pi/m_a$,
where we used the assumption $Q^2/m_a^2 < \epsilon^{-2}$.

Therefore, we find that the energy ( power $P\equiv m_a UV/Q$ ) of the radiation in the space between two flat conductors
is approximately identical to the energy $\sim (\frac{g_{a\gamma\gamma}B_0}{m_a})^2V\rho_aQ^2$ 
( the power $P\sim (g_{a\gamma\gamma}B_0)^2V\rho_aQ/m_a$ ) in resonant cavity with volume $V$.

\section{Summary and Discussion}
We have shown that the strength of the radiation arising from the axion between two flat conductors put parallel to each other
is comparable to the one in resonant cavity. It is never enhanced, although large enhanced electric current
$J_a/(m_a\delta_e)^2$ is expected to flow in the conductors. The expectation
comes from the previous study\cite{1iwazaki} that strongly enhanced radiation is emitted from 
the enhanced electric current in conductor under external
magnetic field. 
The difference is caused by the difference between standing wave arising between two conductors and outgoing wave
from a single conductor. We have clearly shown the difference by taking account of the enhanced electric current in the conductor
and the back reaction of the standing wave on the current. As a consequence, we find that the back reaction
diminishes the electric current and that resultant standing wave is not enhanced.


\vspace{0.1cm}
Finally we show\cite{1iwazaki} how strong the outgoing radiation ( dipole radiation ) from the flat conductor ( slab ) is.
The dipole radiation is coherent so that its flux is proportional to the square $(BS)^2$ of magnetic field $B$ and surface area $S$ of the conductor. 
In particular, we consider ferromagnetic material as a conductor, for instance, neodymium magnet, 
which has strong permanent magnetic field $B\sim 1$T.
Its form is square with surface area $S$. Using the magnet allows us not to prepare apparatus for producing magnetic field.
The magnetic field near the surface is pointed to the identical direction to that of the magnetic field inside the magnet.
For simplicity, we assume that the magnetic field $B$ is parallel ( uniform ) to ( over ) the surface. 
Especially, the magnetic field $\vec{B}_{in}$ inside the magnet is assumed to be parallel ( uniform ) to ( over ) the surface
Then, 
the magnetic field outside the magnet is also parallel to the surface because 
they must be continuous at the boundary. The surface area $A$ we consider
is the area of such a region over which
the magnetic field is uniform.
Then,
the electric field near the surface induced by the axion is $E_a=-g_{a\gamma\gamma}a(t)B$. 
It induces the electric current $J=\sigma E_a$ inside the surface
of the magnet, as we have explained above.  ( It is independent of the permeability $\mu$\,\,\cite{1iwazaki}. )
Noting the skin depth $\delta_e=\sqrt{2/m_a\sigma\mu}$ where $\mu$ denotes recoil
permeability of the magnet, we derive the flux $P$ of the radiation such as $P=m_a^2(S\delta J)^2/3$.  

For instance, using neodymium magnet with magnetic field $B=1$T and surface area $S=100\rm cm\times 100 cm $,
the flux $P$ of the radiation is 

\begin{equation}
P\simeq 1.2\times 10^{-17}\mbox{W}\Big(\frac{m_a}{10^{-4}\mbox{eV}}\Big)
\Big(\frac{\sigma}{4.5\times 10\mbox{eV}}\Big)\Big(\frac{BS}{10000\mbox{Tcm}^2}\Big)^2
\Big(\frac{\rho_a}{0.3\mbox{GeV/cm}^3}\Big)\Big(\frac{g_{\gamma}}{1.0}\Big)^2
\end{equation}
with $g_{\gamma}\simeq 0.37$ for DFSZ model and $g_{\gamma}\simeq -0.96$ for KSVZ model,
where we used $\mu\simeq 1$ and $\sigma \simeq 4.5\times 10$ eV as the parameters of the neodymium. 
The magnetic field is parallel to the surface of the slab.
The radiation is dipole one so that it is mainly emitted to the direction perpendicular to the flat surface of the conductor. 
It is easily detected by using a dish antenna\cite{horn} and broadband receiver over the frequency range $1$GHz to $100$GHz.
Although we need to take into account more realistic configuration of the magnetic field in the magnet,
the estimation indicates a possibility of the use of the magnet for the detection of the dark matter axion.

\vspace{0.2cm}
The author
expresses thanks to Le Hoang Nguyen for useful comments and discussions.
He also expresses great thank to Yasuhiro Kishimoto, Izumi Tsutsui and Osamu Morimatsu for useful comments.
This work is supported in part by Grant-in-Aid for Scientific Research ( KAKENHI ), No.19K03832.



\end{document}